\documentclass[a4paper,12pt]{article}
\usepackage{amsmath}
\usepackage{amssymb}

\begin{document}

\begin{center}
{\bf \Large Symplectic Symmetry of the Neutrino Mass\\[0.2cm]
For Many Neutrino Flavors}\\[0pt]
\vspace{0.5cm}
{\large Nurcan \"Ozt\"urk\footnote{Electronic address:
nurcan@fi.uib.no}}\\[0pt]\vspace{0.5cm}
{\it Department of Physics, University of Bergen\\ All\'{e}gaten 55, Bergen 
N-5007, Norway\\ and\\ 
Department of Engineering Physics, Faculty of Sciences,\\ Ankara University, 
Ankara 06100, Turkey\footnote{Permanent address}}
\end{center}
\vspace{0.3cm}
\begin{abstract}
\par\noindent
The algebraic structure of the neutrino
mass Hamiltonian is presented for two neutrino flavors  considering
both Dirac and Majorana mass terms. It is shown that the algebra is $Sp(8)$ 
and also discussed how the algebraic structure generalizes for the case 
of more than two neutrino flavors.
\vspace{0.2cm}
\par\noindent
PACS: 14.60.Pq, 11.30.Na
\end{abstract}
\vspace{0.2cm}                                    

\section{Introduction}

Recently, the algebraic structure of the most general neutrino mass 
Hamiltonian has been 
discussed by Balantekin and \"Ozt\"urk  \cite{baha}
and it has been found that the algebra is $Sp(4)$ for a single neutrino 
flavor. 
Basically, one can consider the mass term in the Hamiltonian as a 
generalized pairing
problem and for a single neutrino (four Dirac components) the most
general pairing algebra is $SO(8)$ (see, for example, \cite{pere}). 
The authors in [1]
showed that for the neutrino mass Lorentz invariance constrains this
algebra to be $Sp(4)$. 

The Pauli-G\"ursey transformation~\cite{pauli,gursey} which relates a Dirac spinor to its 
charge conjugation corresponds to an $SU(2)$ rotation 
which is 
embedded in the  associated $Sp(4)$ Lie group. A particular Pauli-G\"ursey 
transformation generates the seesaw mechanism~\cite{seesaw} which explains why
neutrino masses are much lighter than those of the quarks and leptons. 
In the literature, the Pauli-G\"ursey transformation plays a crucial role in 
constructing the diquark charges in grand unified theories 
(see, for example,~\cite{saclioglu}). 

It has also been mentioned in \cite{baha} that 
for three neutrino flavors one can introduce three commuting copies of the
$Sp(4)$ algebra and similar arguments follow. It is possible to
introduce a single $Sp(4)$ algebra for three flavors if the individual
masses of the mass eigenstates  are equal up to phases. This case
seems to be too restrictive for model building if one considers 
the recent observations.

The purpose of the present work is to investigate the algebraic structure 
for two neutrino flavors in detail and to  discuss whether the algebraic 
structure gives any constraint on the number of flavors of neutrinos.

\section{The Algebraic Structure For Two Neutrino Flavors}

For simplicity if one starts with 2-neutrino flavors, say $(\nu_e,\nu_\mu)$, 
and considers the Lorentz invariant mass terms; the Dirac mass 
Hamiltonian can be written as \cite{bilenky}
\begin{align}
H_{m}^{D} &= \int  d^{3}{x}\biggl[\bigl(m_{ee}^{D}\bar{\nu}_{eL}\nu_{eR}+h.c.
\bigr)+
\bigl(m_{\mu \mu}^{D}\bar{\nu}_{\mu L}\nu_{\mu R}+h.c.\bigr)\nonumber \\[0.1cm]
&+\bigl(m_{\mu e}^{D}(\bar{\nu}_{\mu L}\nu_{eR}+\bar{\nu}_{eL}\nu_{\mu R})
+h.c.\bigr) \biggr]
\end{align}
and the left- and right-handed Majorana mass Hamiltonian as \cite{bilenky}
\begin{align}
H_{m}^{L} &=\int  d^{3}{x}\biggl[\frac {1}{2}\bigl(m_{ee}^{L}\bar{\nu}_{eL}(\nu
_{eL})^{c}+h.c.\bigr)+\frac {1}{2}\bigl(m_{\mu \mu }^{L}\bar{\nu}_{\mu L}(\nu_{\mu
L})^{c}+h.c.\bigr)\nonumber\\[0.1cm]
&+\bigl(m_{\mu e}^{L}\bar{\nu}_{eL}(\nu_{\mu L})^{c}+h.c.\bigr)\biggr]
\end{align}
\begin{align}
H_{m}^{R} &=\int  d^{3}{x}\biggl[\frac {1}{2}\bigl(m_{ee}^{R}
\bar{\nu}_{eR}(\nu_{eR})^{c}+h.c.\bigr)+\frac {1}{2}
\bigl(m_{\mu \mu }^{R}\bar{\nu}_{\mu R}(\nu_{\mu
R})^{c}+h.c.\bigr)\nonumber\\[0.1cm]
&+\bigl(m_{\mu e}^{R}\bar{\nu}_{eR}(\nu_{\mu R})^{c}+h.c.\bigr)\biggr].
\end{align} 
In these equations $m^D$, $m^L$ and $m^R$ are the Dirac, left- and right-handed
Majorana masses. The last parenthesis in each Hamiltonian represents the mass-mixing terms. As has been done in [1], each mass term can be defined 
as a generator  and the commutations relations between the generators lead 
to a closed algebraic structure. 

Starting with non-mixing terms (the first two 
parenthesis in Eqs.~(1), (2) and (3)) one can write down 
two commuting $Sp(4)$ algebras:
\def\intt{{\textstyle \int}}
\def\half{{\textstyle\frac{1}{2}}}
\def\fourth{{\textstyle\frac{1}{4}}}
\begin{alignat}{2}
D_{+}^{e} &=\intt d^{3}{x}( \bar{\nu}_{eL}\nu_{eR}) & \qquad
D_{+}^{\mu }&=\intt d^{3}{x}( \bar{\nu}_{\mu L}\nu_{\mu R})
\nonumber \\[0.3cm]
D_{-}^{e} &=\intt d^{3}{x}( \bar{\nu}_{eR}\nu_{eL}) & \qquad
D_{-}^{\mu }&=\intt d^{3}{x}( \bar{\nu}_{\mu R}\nu_{\mu L})
\nonumber \\[0.3cm]
D_{0}^{e}&=\half(\nu_{eL}^{\dagger }
\nu_{eL}-\nu_{eR}^{\dagger }\nu_{eR}) & \qquad D_{0}^{\mu }
&=\half(\nu_{\mu L}^{\dagger }
\nu_{\mu L}-\nu_{\mu R}^{\dagger }\nu_{\mu R})\nonumber \\[0.3cm]
L_{+}^{e}&=\half\intt d^{3}{x}[\bar{\nu}_{eL}
(\nu_{eL})^{c}]  & \qquad L_{+}^{\mu }&=\half\intt d^{3}{x}
[\bar{\nu}_{\mu L}(\nu_{\mu L})^{c}]\nonumber \\[0.3cm]
L_{-}^{e}&=\half\intt d^{3}{x}[\overline{(\nu_{eL})^{c}}
\nu_{eL}] & \qquad L_{-}^{\mu }
&=\half\intt d^{3}{x}[\overline{(\nu_{\mu L})^{c}}%
\nu_{\mu L}]\nonumber \\[0.3cm]
L_{0}^{e}&=\fourth\intt d^{3}{x}(\nu_{eL}^{\dagger }\nu_{eL}-\nu
_{eL}\nu_{eL}^{\dagger }) & \qquad L_{0}^{\mu }&=\fourth \intt d^{3}{x}(\nu
_{\mu L}^{\dagger }\nu_{\mu L}-\nu_{\mu L}\nu_{\mu L}^{\dagger })
\nonumber \\[0.3cm]
R_{+}^{e}&=\half\intt d^{3}{x}[\overline{(\nu_{eR})^{c}}
\nu_{eR}] & \qquad R_{+}^{\mu }
&=\half\intt d^{3}{x}[\overline{(\nu_{\mu R})^{c}}%
\nu_{\mu R}] \\[0.3cm]
R_{-}^{e}&=\half\intt d^{3}{x}[\bar{\nu}_{eR}(\nu_{eR})^{c}] 
& \qquad R_{-}^{\mu }
&=\half\intt d^{3}{x}[\bar{\nu}_{\mu R}(\nu_{\mu R})^{c}] 
\nonumber \\[0.3cm]
R_{0}^{e}&=\fourth \intt d^{3}{x}(\nu_{eR}\nu_{eR}^{\dagger }
-\nu_{eR}^{\dagger }\nu_{eR}) & \qquad 
R_{0}^{\mu }&=\fourth \intt d^{3}{x}(\nu_{\mu R}\nu_{\mu R}^{\dagger }
-\nu_{\mu R}^{\dagger }\nu_{\mu R}) 
\nonumber \\[0.3cm]
A_{+}^{e}&=\intt d^{3}{x}( -\nu_{eL}^{T}C\gamma_{0}\nu_{eR}) 
 & \qquad A_{+}^{\mu }
&=\intt d^{3}{x}(-\nu_{\mu L}^{T}C\gamma_{0}\nu_{\mu R})
\nonumber \\[0.3cm]
A_{-}^{e}&=\intt d^{3}{x}(\nu_{eR}^{\dagger}\gamma_{0}C\nu_{eL}^{T}
{}^{\dagger })
 & \qquad 
A_{-}^{\mu }&=\intt d^{3}{x}(\nu_{\mu R}^{\dagger }
\gamma_{0}C\nu_{\mu L}^{T}{}^{\dagger })\nonumber \\[0.3cm]
A_{0}^{e}&=\half\intt d^{3}{x}(\nu_{eL}\nu_{eL}^{\dagger }
-\nu_{eR}^{\dagger }\nu_{eR}) & \qquad A_{0}^{\mu }
&=\half\intt d^{3}{x}(\nu_{\mu L}\nu_{\mu L}^{\dagger }
-\nu_{\mu R}^{\dagger }\nu_{\mu R}) \nonumber
\end{alignat}
The $D$'s, $L$'s, $R$'s and $A$'s in each column satisfy an $SU(2)$ algebra, 
suppressing the $e, \mu$ indices above, one has
\begin{alignat}{3}
[D_{+},D_{-}]&=2D_{0} & \quad, \quad [D_{0},D_{+}]&=D_{+} & \quad, \quad
[D_{0},D_{-}]&=-D_{-}  \\[0.2cm]
[L_{+},L_{-}]&=2L_{0} & \quad, \quad [L_{0},L_{+}]&=L_{+} & \quad, \quad 
[L_{0},L_{-}]&=-L_{-} \\[0.2cm]
[R_{+},R_{-}]&=2R_{0} & \quad, \quad [R_{0},R_{+}]&=R_{+} & \quad, \quad 
[R_{0},R_{-}]&=-R_{-}  \\[0.2cm]
[A_{+},A_{-}]&=2A_{0} & \quad, \quad [A_{0},A_{+}]&=A_{+} & \quad, \quad 
[A_{0},A_{-}]&=-A_{-}
\end{alignat}
and the $D_{0}^{e},A_{0}^{e},D_{0}^{\mu }$ and $A_{0}^{\mu }$ are not 
independent generators:
\begin{alignat}{2}
D_{0}^{e}\equiv L_{0}^{e}+R_{0}^{e} & \quad, \quad D_{0}^{\mu }\equiv L_{0}
^{\mu}+R_{0}^{\mu },\nonumber  \\[0.2cm] 
A_{0}^{e}\equiv R_{0}^{e}-L_{0}^{e} & \quad, \quad A_{0}^{\mu }
\equiv R_{0}^{\mu}-L_{0}^{\mu }.
\end{alignat}
The ten independent generators in each column in Eq.~(4) form an $Sp(4)$ 
algebra, 
say $Sp(4)^e$ and $Sp(4)^\mu$, respectively.

Similarly defining each mixing-mass term as a generator in the following, 
one can easily show that the ten generators $D_{+}^{M},\,D_{-}^{M},\,L_{+}^{M}
,\,L_{-}^{M},\,L_{0}^{M},\,R_{+}^{M},\,R_{-}^{M},\\R_{0}^{M},\,A_{+}^{M},\,A_{-}^{M}$ form another $Sp(4)$ algebra, say $Sp(4)^M$.
\begin{eqnarray}
D_{+}^{M}&=&\intt d^{3}{x}\left( \bar{\nu}_{\mu L}\nu _{eR}+\bar{\nu}_{eL}\nu
_{\mu R}\right) \equiv D_{+}^{1M}+D_{+}^{2M} \nonumber \\[0.3cm]
D_{-}^{M}&=&\intt d^{3}{x}\left( \bar{\nu}_{eR}\nu _{\mu L}+\bar{\nu}_{\mu
R}\nu _{eL}\right) \equiv D_{-}^{1M}+D_{-}^{2M} \nonumber \\[0.3cm]
D_{0}^{M}&=&\half\intt d^{3}{x}(\nu _{eL}^{\dagger }\nu _{eL}+\nu _{\mu
L}^{\dagger }\nu _{\mu L}-\nu _{eR}^{\dagger }\nu _{eR}-\nu _{\mu
R}^{\dagger }\nu _{\mu R})\equiv L_{0}^{M}+R_{0}^{M} \nonumber \\[0.3cm]
L_{+}^{M}&=&\intt d^{3}{x[}\bar{\nu}_{eL}(\nu _{\mu L}{})^{c}] \nonumber \\
[0.3cm]
L_{-}^{M}&=&\intt d^{3}{x[}\overline{(\nu _{\mu L}{})^{c}}\nu _{eL}{}] 
\nonumber \\[0.3cm]
L_{0}^{M}&=&\half\intt d^{3}{x}(\nu _{eL}^{\dagger }\nu _{eL}-\nu _{\mu
L}\nu _{\mu L}^{\dagger })\equiv L_{0}^{e}+L_{0}^{\mu } \\[0.3cm]
R_{+}^{M}&=&\intt d^{3}{x[}\overline{(\nu _{\mu R}{})^{c}}\nu _{eR}{}] 
\nonumber \\[0.3cm]
R_{-}^{M}&=&\intt d^{3}{x[}\bar{\nu}_{eR}(\nu _{\mu R}{})^{c}] \nonumber \\[8pt]
R_{0}^{M}&=&\half\intt d^{3}{x}(-\nu _{eR}^{\dagger }\nu _{eR}+\nu _{\mu
R}\nu _{\mu R}^{\dagger })\equiv R_{0}^{e}+R_{0}^{\mu } \nonumber \\[0.3cm]
A_{+}^{M}&=&\intt d^{3}{x}(-\nu _{eL}^{T}C\gamma _{0}\nu _{eR}-\nu _{\mu
L}^{T}C\gamma _{0}\nu _{\mu R})\equiv A_{+}^{e}+A_{+}^{\mu } \nonumber 
\\[0.3cm]
A_{-}^{M}&=&\intt d^{3}{x(}\nu _{eR}^{\dagger }\gamma _{0}C\nu
_{eL}^{T}{}^{\dagger }+\nu _{\mu R}^{\dagger }\gamma _{0}C\nu _{\mu
L}^{T}{}^{\dagger })^{{}}\equiv A_{-}^{e}+A_{-}^{\mu } \nonumber \\ [0.3cm]
A_{0}^{M}&=&\half\intt d^{3}{x}(\nu _{eL}\nu _{eL}^{\dagger }+\nu _{\mu
L}\nu _{\mu L}^{\dagger }-\nu _{eR}^{\dagger }\nu _{eR}-\nu _{\mu
R}^{\dagger }\nu _{\mu R})\equiv R_{0}^{M}-L_{0}^{M} \nonumber
\end{eqnarray}
The $D^M$'s, $L^M$'s, $R^M$'s and $A^M$'s  satisfy an $SU(2)$ algebra: 
\begin{alignat}{3}
[D_{+}^{M},D_{-}^{M}]&=2D_{0}^{M} & \quad, \quad [D_{0}^{M},
D_{+}^{M}]&=D_{+}^{M}
 & \quad, \quad [D_{0}^{M},D_{-}^{M}]=-D_{-}^{M} \\[0.2cm]
[L_{+}^{M},L_{-}^{M}]&=2L_{0}^{M} & \quad, \quad 
[L_{0}^{M},L_{+}^{M}]&=L_{+}^{M}
 & \quad, \quad [L_{0}^{M},L_{-}^{M}]=-L_{-}^{M} \\[0.2cm]
[R_{+}^{M},R_{-}^{M}]&=2R_{0}^{M} & \quad, \quad 
[R_{0}^{M},R_{+}^{M}]&=R_{+}^{M}
 & \quad, \quad [R_{0}^{M},R_{-}^{M}]=-R_{-}^{M} \\[0.2cm]
[A_{+}^{M},A_{-}^{M}]&=2A_{0}^{M} & \quad, \quad 
[A_{0}^{M},A_{+}^{M}]&=A_{+}^{M}
 & \quad, \quad [A_{0}^{M}, A_{-}^{M}]=-A_{-}^{M} 
\end{alignat}
Note that $D_{0}^{M}$, $L_{0}^{M}$, $R_{0}^{M}$, $A_{0}^{M}$ can be  expressed 
in terms of the other generators as in Eq.~(10).

To find a closed algebraic structure, one has to calculate all the 
commutation relations between the generators of  $Sp(4)^e$, $Sp(4)^\mu$ 
and $Sp(4)^M$. They give the following new generators:

\begin{eqnarray}
N_{+}&=&\intt d^{3}{x(}\nu_{\mu L}^{\dagger }\nu_{eL}-\nu_{eR}^{\dagger
}\nu_{\mu R})\equiv N_{+}^{1}+N_{+}^{2} \nonumber \\[0.3cm]
N_{-}&=&\intt d^{3}{x(}\nu_{eL}^{\dagger }\nu_{\mu L}-\nu_{\mu R}^{\dagger
}\nu_{eR})\equiv N_{-}^{1}+N_{-}^{2} \nonumber \\[0.3cm]
N_0&=&\half\intt d^{3}{x}(\nu_{\mu L}^{\dagger }\nu_{\mu L}-\nu
_{eL}^{\dagger }\nu_{eL}+\nu_{eR}^{\dagger }\nu_{eR}-\nu_{\mu
R}^{\dagger }\nu_{\mu R}) \nonumber \\[0.3cm]
M_{+}&=&\intt d^{3}{x(}\nu_{eL}^{T}\gamma_{0}C\nu_{\mu R}) \nonumber \\[0.3cm]
M_{-}&=&\intt d^{3}{x(-}\nu_{\mu R}^{\dagger }C\gamma_{0}\nu
_{eL}^{T}{}^{\dagger })  \\[0.3cm]
M_{0}&=&\half\intt d^{3}{x}(\nu_{eL}\nu_{eL}^{\dagger }-\nu_{\mu
R}^{\dagger }\nu_{\mu R}) \nonumber \\[0.3cm]
T_{+}&=&\intt d^{3}{x(-}\nu_{eR}^{T}\gamma_{0}C\nu_{\mu L}) 
\nonumber \\[0.3cm]
T_{-}&=&\intt d^{3}{x(-}\nu_{\mu L}^{\dagger }C\gamma_{0}\nu
_{eR}^{T}{}^{\dagger}) \nonumber \\[0.3cm]
T_{0}&=&\half\intt d^{3}{x}(\nu_{\mu L}\nu_{\mu L}^{\dagger }-\nu
_{eR}^{\dagger }\nu_{eR})\nonumber
\end{eqnarray}
The $N$'s, $M$'s and $T$'s satisfy an $SU(2)$ algebra:
\begin{alignat}{3}
[N_{+},N_{-}]&=2N_{0} & \quad, \quad [N_{0},N_{+}]&=N_{+}  & \quad, \quad 
[N_{0},N_{-}]&=-N_{-} \\[0.2cm]
[M_{+},M_{-}]&=2M_{0} & \quad, \quad [M_{0},M_{+}]&=M_{+}  & \quad, \quad 
[M_{0},M_{-}]&=-M_{-} \\[0.2cm]
[T_{+},T_{-}]&=TN_{0} & \quad, \quad [T_{0},T_{+}]&=T_{+}  & \quad, \quad 
[T_{0},T_{-}]&=-T_{-} 
\end{alignat}
also note that $N_{0}\equiv M_{0}- T_{0}$. The commutation relations 
between the  new 
generators in Eq.~(15) and all the others in Eqs.~(4), (10) don't give any other new generator. 
As a result, the 36 independent generators in the following form a Lie
algebra which is the symplectic algebra $Sp(8)$. 
\vspace{8pt}

\qquad$D_{+}^{e},D_{-}^{e},L_{+}^{e},L_{-}^{e},L_{0}^{e},R_{+}^{e},R_{-}^{e},
R_{0}^{e}\qquad$(8 generators)\vspace{8pt}

\qquad $D_{+}^{\mu },D_{-}^{\mu
},L_{+}^{\mu },L_{-}^{\mu },L_{0}^{\mu },R_{+}^{\mu },R_{-}^{\mu
},R_{0}^{\mu }\qquad $(8 generators)\vspace{8pt}

\qquad $D_{+}^{1M},D_{+}^{2M},D_{-}^{1M},D_{-}^{2M},L_{+}^{M},L_{-}^{M},R_{+}^{M},R_{-}^{M},A_{+}^{M},A_{-}^{M}$ \qquad (10 generators)\vspace{8pt}

\qquad $N_{+}^{1},N_{+}^{2},N_{-}^{1},N_{-}^{2}$\ $%
,M_{+},M_{-},M_{0},T_{+},T_{-},T_{0}\qquad $(10 generators)
\vspace{8pt}

In order to illustrate the $Sp(8)$ algebra structure 282 commutation 
relations have been calculated, due to space limitations these relations 
are not given here. 
Fortunately, after presenting the algebraic structure for two neut- rino 
flavors in detail it is straightforward to construct the algebra for 
3, 4, ... neutrino flavors, one obtains $Sp(12)$, $Sp(16)$, ... respectively.
\section{Conclusion}
After having discussed the algebraic structure for a single neutrino flavor 
in~\cite{baha},
it has been shown in the present work that the symplectic algebra structure 
can be extended to many flavors of neutrinos. The detailed calculations 
are given for the two neutrino flavors and the algebra is found to be the 
$Sp(8)$ symplectic algebra. 

For $n$ neutrino flavors the algebraic 
structure can be generalized to $Sp(4n)$. Since one does not obtain a 
non-closed algebra 
by increasing the number of flavors, it may be useful to emphasize that 
the approach based on algebraic structure does not give 
any constraint on the number of neutrino flavors. Nevertheless, it 
would be interesting to investigate if the symplectic algebra 
facilitates the computation of physical neutrino parameters, 
work along this direction is in progress.

\section*{Acknowledgements}

I would like to thank A. Baha Balantekin and A. Ulvi Y{\i}lmazer 
for helpful discussions. I also thank Per Osland for 
the very kind hospitality at the University of Bergen.

\end{document}